\documentclass[twocolumn,showpacs,preprintnumbers,amsmath,amssymb]{revtex4}
\usepackage{amsfonts}
\usepackage{amsmath}
\usepackage{amssymb}
\usepackage{graphicx}
\usepackage{graphics}

\begin{document}

\preprint{submitted to Acta Cryst. D; published in 2005,
\textbf{61}, p.1563}

\title{Enhancing protein drop stability for crystallization by chemical patterning}

\author{Viatcheslav Berejnov}
\email{vb54@cornell.edu}
\author{Robert E. Thorne}
\affiliation{Physics Department, Cornell University, Ithaca, NY,
14853}


\begin{abstract}
Motion of protein drops on crystallization media during routine
handling is a major factor affecting the reproducibility of
crystallization conditions. Drop stability can be enhanced by
chemical patterning to more effectively pin the drop's contact
line. As an example, a hydrophilic area is patterned on an
initially flat hydrophobic glass slide. The drop remains confined
to the hydrophilic area, and the maximum drop size that remains
stable when the slide is rotated to the vertical position
increases. This simple method is readily scalable and has the
potential to significantly improve outcomes in hanging and sitting
drop crystallization.
\end{abstract}

\maketitle

\section{Introduction}
Sitting- and hanging-drop vapor diffusion methods are among the
most popular techniques for protein crystallization (Hampton
Research Co., 2003; McPherson, 1999). In the sitting drop
technique, protein crystals can sediment onto the glass or plastic
surface supporting the drop, frequently adhering so strongly that
they are damaged during retrieval. Adhered crystals are much more
likely to crack during growth because of stresses associated with
contact with the support.

The hanging drop method largely eliminates these problems.
Crystals sediment away from the supporting surface toward the
liquid-air interface (although some may still nucleate and grow on
the supporting surface). Freely suspended crystals often have
unperturbed facets, show less cracking and have smaller
mosaicities, and (in the absence of protein "skins") are easily
retrieved for diffraction studies. Consequently, hanging drops are
preferred for final crystallization trials to obtain the largest,
highest-quality crystals for data collection.

However, obtaining well-formed hanging drops with well-defined
surface-to-volume ratios can be difficult. Depending on mother
liquor composition, the drop can move relatively freely over the
hydrophobic siliconized glass cover slides used to minimize drop
spreading. While the cover slide is being inverted, larger drops
can slide off entirely, and smaller drops can become distended,
changing their shape and rate of equilibration with the well
solution. Similar problems occur for drops dispensed onto polymer
(e.g. teflon) films.

We have demonstrated a simple method to fix the drop shape and
position. Chemical patterning of a circular area increases the
strength of pinning of the drop's contact line, and thereby
defines the drop shape. Drops can be flipped without sliding or
permanently distorting, and the drop area and curvature for given
drop volume can be customized. Furthermore, the drop size and
shape for a given drop volume - which determines the equilibration
time with the well - is now much more reproducible. This
eliminates an important variable factor affecting nucleation rates
and subsequent growth in both hanging and sitting drop vapor
diffusion growth.

\section{Model for Drop Behavior on an Inclined Surface}

The boundary of a liquid drop on the surface of a flat substrate
is a line separating liquid, solid (substrate) and vapor (air),
called the contact line. The angle between the tangent to the
vapor-liquid interface at the contact line and the substrate gives
the contact angle $\theta$ (Figure 1). On a horizontal, perfectly
uniform substrate, the contact angle would have a unique value
determined by the liquid-vapor, liquid-substrate and
vapor-substrate surface energies. If such a substrate were tilted
from the horizontal, the drop would immediately slide off.

On a real substrate, the contact line is pinned by interaction
with disorder at the substrate surface, so that the contact angle
is no longer unique but depends on the drop's history (e.g., how
it was dispensed and how it has been tilted). The strength of the
contact line pinning is reflected in the range of contact angles
that can be supported before the contact line and thus the drop
depin and move.

When a substrate supporting a drop is rotated from the horizontal,
the drop initially distorts while its contact line remains pinned
and fixed, as shown in Figure 1(a,b). As the substrate is rotated
toward the vertical, the net of the gravitational and surface
tension forces may locally exceed the pinning force. The contact
line may then become locally unstable and distort from its
initially circular form, producing a "downhill" spreading of the
drop. Upon further rotation, the gravitational force may exceed
the maximum pinning force on the drop as a whole. The entire drop
then becomes unstable and slides off the substrate.

To understand this process more formally, consider a simplified,
essentially two-dimensional (2D) model shown in Figure 1(b)
(MacDougall and Ockent, 1942; Frenkel, 1948), which can be
extended to three dimensions (Furmidge, 1962; Dussan and Chow,
1983). We assume the drop's contact line is pinned by some
specific but isotropic interaction with the substrate. In this
simplified geometry there are only two contact angles $\theta_a$
and $\theta_r$ defining the angle of the tangent to the air-liquid
interface at the advancing (downhill) and receding (uphill) edges
of the contact line. For a true 3D drop as in Figure 1(a), the
contact angle $\theta$  varies along the (initially circular)
contact line between extreme values at its uphill and downhill
edges (Brown et al.,1980; Rotenberg et al., 1984, El Sherbini and
Jacobi, 2004).

When the substrate is horizontal, $\theta_a=\theta_r$. When the
substrate is tilted from the horizontal by an angle $\alpha$, the
contact line remains pinned, the advancing contact angle
$\theta_a$ increases and the receding contact angle $\theta_r$
decreases. For some critical tilt $\alpha=\alpha^*$ (corresponding
to values of $\theta_a={\theta_a}^*$ and $\theta_r={\theta_r}^*$),
a sufficiently large drop will depin and slide down the substrate.
Balancing the downhill component $F_{g,\shortparallel}$ of the
force due to gravity $F_g$ with the uphill force due to
capillarity $F_c$ provides a boundary condition that must be
satisfied by a stationary pinned drop (Frenkel, 1948). The drop
remains stationary and pinned if for a given inclination, drop
mass, and contact conditions we are able to find a drop shape
consistent with this boundary condition (Frenkel, 1948; Lawal and
Brown, 1982). Beyond a critical inclination angle $\alpha^*$, no
solution can be found and $F_{g,\shortparallel}>F_c$.

The net capillary force is very roughly given by $F_c=l \sigma
\Delta(\cos\theta)$, where $l$ is the characteristic width of the
drop, $\sigma$ is the vapor-liquid surface tension, and
$\Delta(\cos\theta)=\cos\theta_r-\cos\theta_a$. Analogous to a
static friction force, $\Delta(\cos\theta)$ increases as the tilt
angle increases so as to keep the drop stationary. The maximum
value $\Delta(\cos\theta)_{max} =
(\cos\theta_r-\cos\theta_a)_{max}$ that can be sustained by the
contact line pinning is called the contact angle hysteresis (De
Gennes, 1985). For  $\alpha>\alpha^*$, the drop slides and
$\Delta(\cos\theta)$ depends upon the flow conditions (Dussan and
Chow, 1983).

The downhill component of the drop's weight $F_g$ is
$F_{g,\shortparallel} = \rho g V \sin\alpha$, where $\rho$ is the
density of the liquid, $g$ is acceleration due to gravity, and $V$
is the drop volume. For the 2D drop of Figure 1(b), $V\sim alh$
where $a\sim l$ is the drop's diameter and $h$ is its
characteristic height. Combining the formulas for $F_c$ and
$F_{g,\shortparallel}$ yields the condition for drop stability
$F_{g,\shortparallel} = F_c$ on a surface inclined at angle
$\alpha$:
\begin{equation}
\rho g a^2 h \sin\alpha= l \sigma
(\cos\theta_r-\cos\theta_a)\label{eq:eq1}
\end{equation}
For any angle $\alpha$ between 0 and $\alpha^*$ the drop is able
to adjust its surface shape by varying the contact angle values
$\theta_a$ and $\theta_r$ and thus $\Delta(\cos\theta)$. The last
contact angles $\theta_a^*$ and $\theta_r^*$ the drop reaches at
$\alpha^*$ are determined by the contact angle hysteresis
$\Delta(\cos\theta)_{max}$. At that critical point,
$(\cos\theta_r-\cos\theta_a)$ corresponds to a surface energy
difference that is an invariant of the drop shape.

For hanging drop crystallization, we are particularly interested
in the case $\alpha=90^\circ$ when the drop is oriented
vertically. For such a drop to be stable against sliding,
$V_c(\sin\alpha=1)=const\approx\frac{l \sigma}{\rho g}
\Delta(\cos\theta)$. A drop satisfying this criterion is
absolutely stable. It may still move, however, if accelerations
during flipping or other handling - which produce forces on the
drop that add to the gravitational force - are large enough. The
stability criterion depends on the properties of the drop, which
are affected by the presence of protein.  Proteins have
hydrophobic and hydrophilic parts and thus can behave as
surfactants (Mobius and Miller, 1998). Protein present at the
vapor/liquid interface decreases the solution's surface tension.
In addition, protein adsorption on the substrate surface can
change the contact angles, surface pinning properties, and contact
angle hysteresis.

In protein crystallization, we often want to minimize the drop
surface-to-volume ratio to produce slow equilibration with well
solutions favorable for nucleation. For hanging drop
crystallization, we would also like to maximize the stability of
the drop, equivalent to obtaining the largest possible drop volume
that is stable at $\alpha=90^\circ$. Increasing the drop volume
will decrease the drop surface-to-volume ratio until drop
flattening by gravity becomes important. Increasing the contact
angle hysteresis (i.e., the strength of contact line pinning) will
increase the maximum size of an absolutely stable drop.

By chemically patterning a substrate to produce hydrophobic and
hydrophilic areas so as to increase the strength of contact line
pinning and $\Delta(\cos\theta)_{max}$, we show in the following
that the maximum absolutely stable drop volume can be increased.
Strong pinning of the contact line at the hydrophobic/hydrophilic
boundary also yields much more accurate control over drop shape
and surface-to-volume ratio.

\section{Materials and Methods}
Six times recrystallized and lyophilized hen egg white lysozyme
from Seikagaku America (Falmouth, MA) was dissolved in a sodium
acetate (NaAc) aqueous buffer, prepared by adding concentrated
acetic acid to a 50 mM sodium acetate solution to adjust the pH to
5.0, where M=[mol/L]. Protein concentration was measured using a
Spectronic Genesys TM 5 spectrophotometer (Spectronic Instrument,
NY).

Solution surface tension as a function of protein concentration
was measured using the pendant drop counting method. A pipette tip
is filled with protein solution and then held vertically, and the
number of drops that fall from the tip in a given time and the
total mass of these drops is measured. Using these values, the
drop radius as it emerges from the tip, and the liquid density,
the vapor-liquid surface tension can be calculated. For the
protein solutions studied the drop radius was measured to be
roughly $11\%$ larger than the external tip diameter.

Siliconized and initially flat glass slides HR3-231 with diameter
22 mm were purchased from Hampton Research (Laguna Niguel, CA). On
a freshly unpackaged slide, a 40 $\mu$l buffer drop formed a
reproducible contact angle of $90^\circ-92^\circ$.

To increase contact angle hysteresis and thus drop stability under
inclination, a hydrophilic circular region was fabricated on some
of these hydrophobic siliconized glass slides. A drop of 1 M NaOH
was placed on the surface at room temperature, and then the slide
was heated on a hot plate in air to $100^\circ$C until all liquid
evaporated. The base removes the silanol layer and reacts with the
glass, producing water-soluble silicates. These were washed away
using a jet of distilled deionized water, and the slide then dried
using pure $N_2$ gas. NaOH drop volumes of 25$\mu$l, 50$\mu$l,
75$\mu$l, 100$\mu$l were used to produce etched areas with
diameters of $5.5\pm0.5$mm, $7.0\pm0.5$mm, $8.0\pm.5$mm, and
$9.5\pm0.5$mm. Following this treatment, the etched areas were
highly hydrophilic, with contact angles for pure water of less
than $1^\circ$. When a drop was dispensed onto the etched region,
the large difference in contact angle between the etched and
unetched regions strongly pinned the drop's contact line at the
boundary between these regions. This strong pinning produced a
contact angle hysteresis for pure water of $90\pm1^\circ$,
compared with value of $12\pm1^\circ$ for a uniform siliconized
glass slide.

The patterned and unpatterned slide surfaces were analysed by
contact mode AFM (DI MultiMode III, Santa Barbara, CA) using NSC
1215 tips from MikroMasch. A typical surface profile in Figure 3
shows that the treated area is etched to a depth of about 1
$\mu$m. To determine drop stability on a given unpatterned or
patterned slide, a drop was dispensed onto a horizontal slide
using a 100 $\mu$l micropipette (Pipetman Co., France). The slide
was then slowly rotated in $2^\circ-4^\circ$ steps, allowing
roughly one minute after each rotation to allow transient
relaxations of the drop to die out. The contact angles of the drop
were measured using a vertical goniometer, and its shape and
position recorded using a digital camera. Dispensed drop volumes
were accurate to $1\%$, and tilt and contact angle measurements
were accurate to $1^\circ-2^\circ$. Each measurement on an
"unpatterned slide" used a fresh slide, and each measurement on a
"patterned slide" used a new fabricated slide, to avoid surface
contamination problems between measurements.

\section{Results and Discussion}
Drop stability on patterned and unpatterned slides was
investigated as a function of drop orientation angle, drop volume,
drop area, and protein concentration. As the inclination angle was
incremented upward, the contact angles and contact line position
varied around the drop circumference, in part through local
instabilities (similar to avalanches) that caused abrupt jumps in
contact line position. For larger drops, the drop became
absolutely unstable at a critical angle $a^*\leq 90^\circ$ and
slid off the slide, as shown in Figures 2 and 4. To construct a
drop stability phase diagram, the patterned area (radius) and
protein concentration were fixed, and the critical angle measured
as a function of drop volume. Repeating these measurements with
different protein concentrations and patterned areas provided a
complete characterization of drop stability.

Figure 4 shows a typical drop stability diagram for patterned and
unpatterned slides with fixed patterned radius and fixed protein
concentration.  For a given inclination $\sin\alpha$, a drop
remains stable (even though its contact line may undergo local
displacements around the drop's circumference) for volumes up to a
critical volume $V_c(\sin\alpha)$. Drops smaller than the critical
volume $V_c(\sin\alpha=1)=const$ will be absolutely stable; they
will not slide off even at $\alpha=90^\circ$. Drops larger than
$V_c$ become unstable and slide off at a critical inclination
$\sin\alpha_c\sim 1/V$ as follows from Equation 1. For the
conditions in Figure 4, substrate patterning increases the volume
range of absolute stability by $\sim15\%$.

For hanging drop crystallization, we want to maximize the range of
volumes for which a drop is absolutely stable and thus the
critical volume $V_c(\sin\alpha=1)$. $V_c$ depends on the protein
concentration and on the area of the patterned region, which
determines the drop's area.

Figure 5 shows how the critical volume for absolute stability
$V_c(\sin\alpha=1)$ varies with protein concentration. Drop
volumes below each curve are absolutely stable. For both
unpatterned and patterned slides, the critical volume decreases
beyond a protein concentration of roughly 1 mg/ml. This behavior
is consistent with the measured increase in drop surface tension
with increasing protein concentration (Mobius and Miller, 1998).

For pure, protein-free buffer on an unpatterned slide, the
critical volume $V_c$ is 12.5 $\mu$l, a factor of 3.2 smaller than
for the smallest protein concentration (0.01 mg/ml) shown by solid
circles in Figure 5.  Thus, variation of $V_c$ with protein
concentration on unpatterned slides is non-monotonic. This complex
behaviour likely results because protein adsorbs to the slide
surface, modifying the contact angle and contact angle hysteresis.
Even very small solution concentrations of protein are sufficient
to modify these properties. On the other hand, the critical volume
$V_c$ for the patterned slides (open circles) is constant between
0 mg/ml and 0.01 mg/ml, indicating that contact line pinning
dominates the drop-substrate interaction.

Over the whole concentration range, patterning provides large
increases in the critical volume for absolute stability, from
$45\%$ at C=0.01 mg/ml to $48\%$ at 100 mg/ml. At concentrations
that yield the largest protein crystals (near 2 mg/ml), the
increase is $38\%$.

Figure 6 shows how the critical volume $V_c(\sin\alpha=1)$ for
absolute stability varies with the drop diameter $D$ (measured
when the slide is horizontal) for pure buffer and for solutions
with different protein concentrations. To simplify representation
of the data, Figure 6 uses dimensionless variables
$\widetilde{V_c}$ and $\widetilde{D}$ defined by the formulas
$\widetilde{V_c}=(V_c-V_{c0})/V_{c0}$ and
$\widetilde{D_c}=(D_c-D_{c0})/D_{c0}$, respectively.   Here
$V_{c0}=12.5$ $\mu$l and $D_0=3.5$ mm are the critical volume and
corresponding wetting diameter of the drop measured for
protein-free buffer solution on an unpatterned, hydrophobic
siliconized slide, indicated by point A.

Points B and C are also for unpatterned slides, but with protein
concentrations of 1 mg/ml and 100 mg/ml, respectively.  At 1 mg/ml
(point B), the critical volume is 41.4 $\mu$l, 3.3 times larger
than for protein-free drops.  Increasing the protein concentration
to 100 mg/ml (point C) decreases the critical volume to 32.5
$\mu$l, 2.6 times larger than for protein-free drops.

On patterned slides, the wetted drop diameter is equal to the
patterned diameter, and can be made much larger than on
unpatterned slides. For protein-free buffer (solid circles in
Figure 6), increasing the wetted diameter from 3.5 mm to 9.5 mm
increases the critical volume by a factor of 4.6.  For 1 mg/ml and
100 mg/ml protein, increasing the diameter from 6.1 and 5.1 mm,
respectively, to 9.5 mm increases the critical volume by roughly
$20\%-25\%$. This increase comes at the cost of a factor of 0.94
decrease in drop surface to volume ratio.

\section{Conclusions}
We have demonstrated that the overall stability of drops on an
inclined flat substrate can be significantly increased by
patterning the substrate surface so as to strongly pin the drop's
contact line.  This allows larger drops to be used in hanging drop
crystallization.    More important, patterning precisely defines
the drop shape and thus its surface-to-volume ratio, independent
of how the drop is dispensed and flipped. Patterned substrates
should thus provide more repeatable drop equilibration rates with
well solutions and thus more reproducible crystal nucleation and
growth.

The largest increases in drop stability and volume occur for
protein-free solutions.  This suggests that surface patterning can
be used to define "well" drops on the same substrate as the
protein drop, and with volumes 10-100 times the protein drop
volume.  Consequently, with surface patterning wells can be
eliminated, yielding a crystallization platform that can be used
in any orientation.

\begin{acknowledgments}
This work was funded by the NASA (NAG8-1831) and by the NIH (R01
GM65981-01).
\end{acknowledgments}

\section{FIGURE CAPTIONS}

Figure 1. \\Photograph of a protein-containing drop on a
siliconized glass slide. (b) Simplified 2D model for a pinned drop
on a flat inclined substrate.\\

Figure 2.\\
Effect of substrate patterning on stability of protein-free buffer
drops.  The numbers in each frame indicate the tilt angle $\alpha$
in degrees. Substrates are hydrophobic siliconized glass slides.
Patterned slides have circular hydrophilic areas with diameters of
8 mm. \\

Figure 3.\\
AFM image of the boundary between untreated and NaOH-treated parts
of the slide. The average etched depth of the NaOH-treated
hydrophilic area is about 1 $\mu$m.\\

Figure 4.\\
Stability diagram for drops on unpatterned siliconized slides
(open squares) and siliconized slides patterned with an 8 mm
diameter hydrophobic region (solid squares), for drops containing
1 mg/ml lysozyme in 50mM NaAc buffer at pH=5. The critical volumes
beyond which drops are unstable at $\alpha=90^\circ$ are
$V_c=47\pm2$ $\mu$l and $V_{c,p}=55\pm2$ $\mu$l for the
unpatterned and patterned slides, respectively. Drops with smaller
volumes are absolutely stable. Dashed lines represent the function
$\sin\alpha\sim1/V$. \\

Figure 5.\\
Critical drop volume $V_c(\sin\alpha=1)$ for stability on
vertically oriented substrates versus protein concentration. The
open and solid circles show data for unpatterned siliconized
slides and siliconized slides patterned with a 9.5 mm diameter
hydrophobic region, respectively. The approximate concentration
range over which crystal growth can be induced by addition of NaCl
precipitant is indicated.\\

Figure 6.\\
Enhancement of drop critical volume $\widetilde{V_c}$ by
patterning a circular hydrophilic region of diameter
$\widetilde{D}$. The points A, B, and C correspond to drops on
unpatterned slides. The raw data is scaled by the value of the
point A ($V_{c0}$ = 12.5 $\mu$l and $D_0$ = 3.5mm) as described in
the text. Solid circles correspond to protein free buffer
solution, and open circles and squares correspond to protein
concentrations of 1mg/ml and 100mg/ml, respectively.

\end{document}